\documentclass[letterpaper,twocolumn,english,showpacs,preprintnumbers,amsmath,amssymb]{revtex4-1}
\usepackage{amsmath}
\usepackage{amssymb}
\usepackage{graphicx}
\usepackage{esint}

\makeatletter

\providecommand{\tabularnewline}{\\}

 \@ifundefined{textcolor}{}
 {%
   \definecolor{BLACK}{gray}{0}
   \definecolor{WHITE}{gray}{1}
   \definecolor{RED}{rgb}{1,0,0}
   \definecolor{GREEN}{rgb}{0,1,0}
   \definecolor{BLUE}{rgb}{0,0,1}
   \definecolor{CYAN}{cmyk}{1,0,0,0}
   \definecolor{MAGENTA}{cmyk}{0,1,0,0}
   \definecolor{YELLOW}{cmyk}{0,0,1,0}
 }

\usepackage{color}\usepackage{epsfig}\usepackage{subfigure}\usepackage{dcolumn}\usepackage{stmaryrd}\usepackage{mathrsfs}\usepackage{pifont}\usepackage{amsthm}\usepackage{bm}\usepackage{latexsym}\usepackage{amsfonts}

\newcommand{\beq}{\begin{equation}}
\newcommand{\eeq}{\end{equation}}
\newcommand{\beqa}{\begin{eqnarray}}
\newcommand{\eeqa}{\end{eqnarray}}

\usepackage{babel}

\makeatother

\usepackage{babel}
\begin{document}

\title{Environment-Assisted Quantum State Restoration Via Weak Measurement}

\author{Kelvin Wang$^{1,2}$}

\email{kelvinwang42@gmail.com}

\author{Xinyu Zhao$^{1}$}


\author{Ting Yu$^{1}$}


\affiliation{$^{1}$Center for Controlled Quantum Systems and the Department of
Physics and Engineering Physics, Stevens Institute of Technology,
Hoboken, New Jersey 07030, USA.}

\affiliation{$^{2}$Academy for the Advancement of Science Technology at Bergen
County Academies, Hackensack, New Jersey, 07601, USA.}

\date{\today}
\begin{abstract}
In this paper, a new quantum state restoration scheme is proposed based on
the environment-assisted error correction (EAEC) scheme.  By introducing
a weak measurement reversal (WMR) operation, we shall show how to
recover an initial state of a quantum  open system without invoking random unitary decompositions which are known to be absent in many important
physical systems. We illustrate 
our new scheme and compare it with a scheme purely based on the WMR operation in the case of dissipative channel.  
In the proposed new scheme,  the successful probability of recovering an unknown initial state can be significantly improved 
when the information obtained from an environment measurement is taken into account.  Moreover, 
 we show that the applicable range of our proposed scheme 
is  wider than the EAEC scheme. Finally, the optimization of the 
successful probability for possible Kraus decompositions is discussed.

\end{abstract}


\maketitle

\section{\label{sec:I}Introduction}

Decoherence is  a major obstacle in effectively  implementing a quantum information processing task 
due to the inevitable interactions between quantum systems and their environments
 \cite{Book1,Book2,QCQI,WW,XX,YY,Yu-Eberly04,Nori}.   In the past few decades, several 
theoretical schemes, including dynamic decoupling control \cite{DD}, feedback control \cite{FBC}, error correction
codes \cite{ECC},   weak measurement reversal (WMR) \cite{WK,Kim} and
environment-assisted error correction (EAEC) \cite{Strunz,LostFound}, to name a few, have been proposed to protect a quantum 
state from environmental noises. 
The EAEC scheme originally proposed by
Gregoratti and Werner \cite{LostFound} has been studied in several interesting contexts
\cite{Buscemi1,Buscemi2,EC_Depolarizing,EC_dissipative,Exp_ECC,Strunz,Lidar2010,Rosgen2008,Hayden2005,Audenaert2008}.
The underlying crucial idea in the EAEC scheme is to restore an unknown quantum state coupled to a noisy environment
by performing a measurement on the environment followed  by
a quantum restoration operation on the system of interest conditioned 
on environment measurement outputs.  Essentially,  if the non-unitary evolution of a quantum open system can 
be randomly decomposed into many unitary quantum processes, then the initial state can be recovered for each particular realization.   
It should be noted that, in the original Gregoratti-Werner  scheme,  in order to restore the initial system state,  only one
reversal operation conditioned on the outcome of a
measurement performed on the environment  is needed.  In addition,  it is shown that the successful probability is 
always $1$, and the fidelity of the recovered state is also always $1$.  While the EAEC scheme is clearly  of interest  
in quantum decoherence control and quantum state restoration, its application is restricted due to the nonexistence of 
the required random unitary (RU) decomposition for many realistic physical models.  
A recent  attempt in relaxing the RU limitation is to use  a non-RU decomposition to recover some entangled states 
in a dephasing channel \cite{Zhao2013}.   However, a general scheme
dealing with an arbitrary unknown state and a noisy  channel that does not allow an RU decompositions 
is still an open question.

The dynamics of  a quantum open system, represented by a reduced density operator $\rho(t)$,  can be formally 
described by a Kraus decomposition \cite{Choi,Lind,Kraus,QCQI}, where $\rho(t)={\textstyle \sum_{n}}K_{n}(t)\rho(0)K_{n}^{\dagger}(t)$.
The crucial point of implementing the EAEC scheme is that all the Kraus
operators $K_{n}$ must be proportional to unitary operators, i.e., $K_{n}=c_{n}U_{n}$. If this is satisfied, then these $K_n$ operators are said to be a 
random unitary (RU) decomposition. If a Kraus operator $K_{n}$
is not an RU type, the Gregoratti-Werner  scheme cannot be used to recover an arbitrary initial state as the reversal quantum
operations are not available in this case.  We shall show, however, that combing environment measurement and
weak measurement \cite{WK,Kim} can yield a useful technique to restore a quantum state in a dissipative channel.
Unlike the RU decomposition, the successful probability of the new scheme cannot be 1.
 We shall consider a case  where  $K_{n}$
is not promotional to unitary matrix,  but still a non-singular matrix.  In this case, as shown later,  by using
weak measurement described by a positive-operator valued measure (POVM). 
 it is still possible to find a reversal operation to restore the post-measurement state $K_{n}\rho(0)K_{n}^{\dagger}$.
As a result, the new scheme we propose here  is no longer a deterministic process.  However, the advantage of this new scheme is its wider  range of applications
and the high fidelity of the recovering operations.

This paper is organized as follows: In Sec.~\ref{sec:II}, we shall briefly
review the EAEC scheme. Then, we propose a new version of the EAEC scheme
with the aide of weak measurement reversal operations in Sec.~\ref{sec:III}. In
order to illustrate the new scheme, we analyze carefully a two-qubit
dissipative model and compare the results with
the  WMR scheme in Sec.~\ref{sec:IV}.  In Sec.~\ref{sec:V},  we discuss how the Kraus decomposition 
affects the successful probability of the newly proposed scheme.
Finally, a conclusion and the future perspective are given in Sec.~\ref{sec:VI}.

\section{\label{sec:II} Brief overview of environment-assisted error correction
scheme}

The dynamics of an open quantum system may be established from
the quantum dynamics describing the total system (system plus environment),
which is governed by the Schr\"{o}dinger equation. The evolution for
the reduced density matrix of the system of interest can be obtained by tracing
over the environmental degrees of freedom $\rho(t)=\text{tr}_{E}[\rho_{{\rm tot}}(t)]$,
and it can be  formally written in the Kraus (operator sum) representation
\cite{Choi,Kraus,QCQI,Lind},
\beqa
\rho(t)&=&{\textstyle \sum_{n}}\langle \psi_n|U|0\rangle\rho(0)\langle0|U^{\dagger}|\psi_n\rangle\nonumber \\
 &=& {\textstyle \sum_{n}}K_{n}\rho(0)K_{n}^{\dagger},\label{Kraus}
\eeqa
where $K_{n}=\langle \psi_n|U|0\rangle$ are the zero-temperatrue ($T=0$) Kraus operators,
which depend on the initial state of the environment and the choice
of the complete basis of the environment $\{|\psi_n\rangle\}$. 

The EAEC scheme is based
on the idea that the environment will collapse into an eigenstate of the measured observable if
a measurement on the environment is performed. Correspondingly, after the measurement is performed, 
the system will be projected into a state relative to each resulting environmental state. More specifically,  if the $n^{\text{th}}$
outcome in the  environment measurement is observed then the system
will be in the state represented by $\rho_{n}(t)=K_{n}\rho(0)K_{n}^{\dagger}$
(up to a normalization factor).  Thus, the EAEC scheme is implementable if the Kraus decomposition 
is random unitary (RU), i.e., $K_{n}=c_{n}U_{n}$ for each $n$, where
$U_{n}^{\dagger}=U_{n}^{-1}$ and $c_{n}$ satisfy $\sum_{n}|c_{n}|^{2}=1$. The 
initial quantum state can then be recovered by applying a quantum reversal operation,
\begin{equation}
\rho_{R}(t)=R_{n}\rho_{n}(t)R_{n}^{\dagger}=\rho(0),\label{eq:3}
\end{equation}
where $R_{n}=\frac{1}{c_{n}}U_{n}^{-1}$. It is clear that if there exists an RU decomposition, 
the fidelity and probability of realizing restoration are always one.  But it should be noted that 
that the desired RU decomposition is not always available. For example,  the RU decomposition does not exist 
for a dissipative channel.

Therefore, it is of interest to extend the original EAEC to a probabilistic quantum reversal scheme
where the RU decomposition is not required.  We shall show in what follows that, for
some open systems, the non-unitary quantum evolution may  be reversed by
a quantum weak measurement reversal (WMR) operation \cite{WK,Kim}. The quantum weak
measurement can be represented by a set of generalized measurement
operators $\{M_{m}\}$ satisfying
\begin{equation}
\sum_{m}M_{m}M_{m}^{\dagger}=I.\label{normalization}
\end{equation}
All of these operators $M_{m}M_{m}^{\dagger}$ must be positive semidefinite,
so that it is called positive-operator valued measure  (POVM).
When a POVM type measurement is performed, the post-measurement state becomes
\begin{equation}
\rho^{\prime}=\frac{M_{m}\rho M_{m}^{\dagger}}{{\rm tr}(M_{m}\rho M_{m}^{\dagger})},
\end{equation}
assuming that the $m^{th}$ outcome is observed in the weak measurement. When the measurement operators are chosen as a complete set of orthogonal projectors, $M_{m}M_{m}^{\dagger}=|\phi_{m}\rangle\langle\phi_{m}|$, the generalized POVM reduces to a projective (strong) measurement.

The weak measurement can be implemented in many different ways depending on the concrete physical systems. For example, in an optical system, the qubit states are encoded by the polarization of light. Using a Brewster-angle glass plate \cite{Kim}, the vertical (or horizontal) polarization component is partially blocked. In another example, by introducing an auxiliary qubit, the weak measurement on the system qubit can be achieved through a strong (projective) measurement on the auxiliary qubit with some necessary quantum gate operations \cite{WK}.

In the next section, we will show that an unknown initial state may be restored using a probabilistic weak measurement restoration 
operation even if the Kraus operators are not in an RU form.

\section{\label{sec:III} Environment-assisted error correction with weak measurement
reversal}

Our proposed scheme is a combination of a weak measurement operation
and a measurement on the environment.  In this new scheme, we only need
to assume that some of the Kraus operators $K_{n}$ in the decomposition Eq.~(\ref{Kraus})
are invertible instead of unitary. That is, the inverse operator $K_{n}^{-1}$
exists, which is denoted as $R_{n}=K_{n}^{-1}$. Then, it is straightforward
to check the following operation will recover the initial states
\begin{equation}
R_{n}K_{n}\rho(0)K_{n}^{\dagger}R_{n}^{\dagger}=\rho(0).
\end{equation}
However, when $R_{n}$ is not a unitary quantum operation, the
implementation of the reversal quantum operation is not straightforward.
Actually, such an $R_{n}$ (unnormalized) operation can be realized
through a weak measurement.  Suppose that there is only one invertible $K_n$ 
operator in the decomposition Eq.~(\ref{Kraus}), the case involving more than one
invertible $K_n$ can be treated in a similar way.  Without loss of generality, we can always choose $R_{n}=K_n^{-1}$ (which is invertible) 
as one of the measurement operators in the complete set $\{M_{m}\}$,
e.g.  $M_{1}\varpropto R_{n}$. The remaining measurement operators $\{M_{m}\}$ ($m\neq1$) in the complete set  can be 
chosen arbitrarily as long as all the  measurement operators form a complete set $\{M_{m}\}$. 
When the measurement on the environment is performed,  the system of interest has a certain probability to collapse into the state
represented by $K_{n}\rho(0)K_{n}^{\dagger}$ ($K_n$ is invertible),  Then the weak measurement will be performed on this state.  
The measurement result represented by $M_1$ can be used to recover to the initial state. Of course, the other measurement results by  $\{M_{m}\}$ ($m\neq1$) will be discarded.

More systematically, we can implement our new ``environment-assisted error correction with weak measurement
reversal'' (EA-WMR) scheme as follows: First, we send an unknown initial
state $\rho(0)$ into a noisy channel. After a period of
time $t$, we make a measurement on the environment of the channel.
For each measurement outcome ``$n$'', the state of the system
correspondingly collapses into the state $K_{n}\rho(0)K_{n}^{\dagger}$. If $K_n$ is invertible, then
a WMR operation $R_{n}=K_{n}^{-1}$ may be used to restore the initial
state as $R_{n}K_{n}\rho(0)K_{n}^{\dagger}R_{n}^{\dagger}=\rho(0)$.   In the WMR operation, we perform a weak measurement $\{M_{m}\}$, satisfying the conditions imposed in the previous paragraph, on the collapsed state $K_{n}\rho(0)K_{n}^{\dagger}$. Then, the initial state can be restored with certain probability. Note that if $K_n$ is singular (non-invertible) or the weak measurement does not yield the desirable result even if $K_n$ is invertible, then the whole process has to be repeated.  
The procedure is explicitly described in Fig. ~\ref{Sketch}. The differences
between our new scheme and the previous schemes are illustrated in table
\ref{table1}.

\begin{widetext}
\begin{center}
\begin{table} [h]
\noindent \begin{centering}
\begin{tabular}{|c|c|c|c|}
\hline 
Procedure  & Step 1  & Step 2  & Step 3\tabularnewline
\hline 
\hline 
EAEC  & Sending the state to channel  & Measurement on environment  & Restoration (unitary)\tabularnewline
\hline 
WMR$^{1}$  & Weak measurement  & Sending the state to channel  & Restoration (WM)\tabularnewline
\hline 
EA-WMR  & Sending the state to channel  & Measurement on environment  & Restoration (WM)\tabularnewline
\hline 
\end{tabular}
\par\end{centering}

\caption{Differences among EAEC, WMR, and EA-WMR. $^{1}$There are many WMR
schemes, the one listed in the table mainly describe a typical scheme
discussed in Ref. ~\cite{Kim}.}

\label{table1}
\end{table}

\par\end{center}
\end{widetext}

\noindent \begin{center}
\begin{figure}
\noindent \begin{centering}
\includegraphics[width=1\columnwidth]{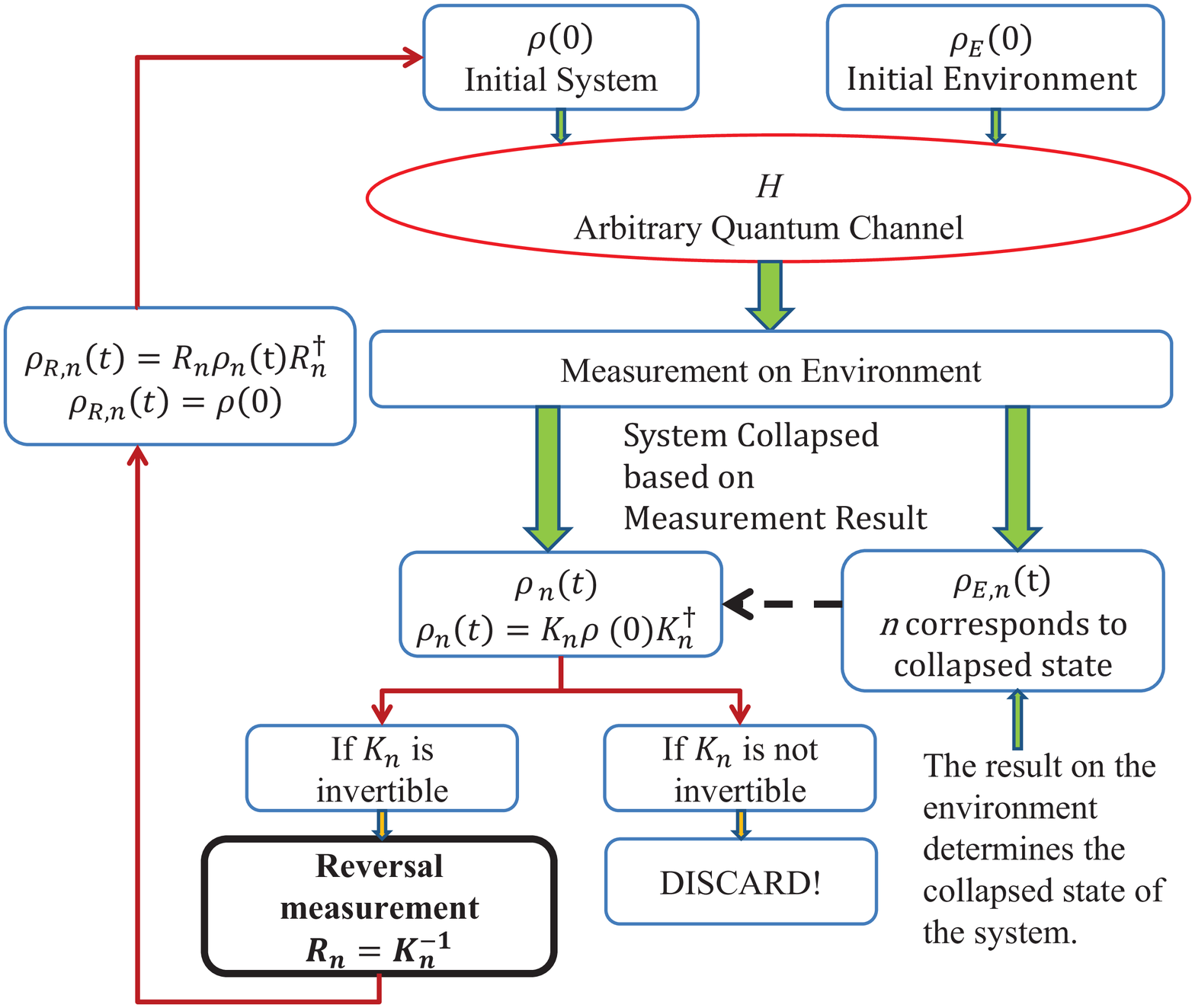} 
\par\end{centering}

\caption{(Color online) Sketch of the EA-WMR scheme.}

\label{Sketch} 
\end{figure}

\par\end{center}

Unlike the original EAEC scheme, which is deterministic, our new scheme
is essentially a probabilistic process arising from the measurement
on the environment and the weak measurement reversal operation. For the
measurement on the environment, the probability of finding the $nth$
outcome is given by, 
\begin{equation}
P_{1n}={\rm tr}[K_{n}\rho(0)K_{n}^{\dagger}].
\end{equation}
In order to compute the successful probability of the subsequent reversal
operation $R_{n}$, we need to use the normalized operator 
\begin{equation}
R_{n}^{\prime}=N_{n}R_{n},
\end{equation}
instead of $R_{n}$, where $N_{n}$ is given by 
\begin{equation}
N_{n}=\min\{\sqrt{\lambda_{i}},i=1,2,3,\dots\},\label{eqN}
\end{equation}
where $\lambda_{i}$ are the eigenvalues of the matrix $K_{n}K_{n}^{\dagger}$.
Note that physically $R_{n}$ and $R_{n}^{\prime}$ represent the
same operation, while the normalization requirement purely
arises from the mathematical convention that the total probability
is $1$ as shown in Eq.~(\ref{normalization}). 
Therefore, the operators $R_{n}^{\prime}$ will
be used when we compute the probability.  For an environment-measurement
outcome ``$n$'', the successful probability for the subsequent
reversal operation is given by 
\begin{equation}
P_{2n}=\frac{{\rm tr}[R_{n}^{\prime}K_{n}\rho(0)K_{n}^{\dagger}R_{n}^{\prime\dagger}]}{{\rm tr}[K_{n}\rho(0)K_{n}^{\dagger}]}.
\end{equation}
Then, the final successful probability for our new scheme is
\cite{Normalization} 
\begin{equation}
P_{EW}=\sum_{n}P_{1n}P_{2n}={\rm tr}[R_{n}^{\prime}K_{n}\rho(0)K_{n}^{\dagger}R_{n}^{\prime\dagger}]=\sum_{n}N_{n}^{2}.\label{PEW}
\end{equation}
Now, it is clear that the successful probability of recovering the
initial state crucially depends on the Kraus decomposition. Therefore, it is interesting to 
consider how to improve the probability by choosing a ``better'' Kraus decomposition. The optimization of the Kraus
decomposition will be discussed in Sec.~\ref{sec:V}.  Before that,
we will compare our new scheme with the WMR scheme in a dissipative
model.

\section{\label{sec:IV} Successful probability for WMR and EA-WMR for two qubits: Dissipative model}

First, let us point out that the original EAEC can not be applied to the dissipative model due to
the absence of a viable RU decomposition.   
In the following example, we consider two qubits coupled to two individual baths at zero temperature ($T=0$), in order to make a comparison to Ref. \cite{Kim}, where it is shown that  the weak measurement reversal can be used to recover the initial state in a probabilistic manner. It should be noted that the finite temperature case can be treated in a similar manner \cite{Wang}.  Now we compare the newly designed EA-WMR scheme with the WMR scheme.  For the two-qubit dissipative model,  the Kraus operators are given by, 
\begin{equation}
\rho(t)=\sum_{i=1}^{4}K_{i}\rho(0)K_{i}^{\dagger},
\end{equation}
where the Kraus operators $K_{i}$ can be explicitly written as \cite{Yu-Eberly04},
\begin{equation}
K_{1}=\left[\begin{array}{cc}
\gamma_{A} & 0\\
0 & 1
\end{array}\right]\otimes\left[\begin{array}{cc}
\gamma_{B} & 0\\
0 & 1
\end{array}\right],\label{K1}
\end{equation}
\begin{equation}
K_{2}=\left[\begin{array}{cc}
\gamma_{A} & 0\\
0 & 1
\end{array}\right]\otimes\left[\begin{array}{cc}
0 & 0\\
\omega_{B} & 0
\end{array}\right],
\end{equation}
\begin{equation}
K_{3}=\left[\begin{array}{cc}
0 & 0\\
\omega_{A} & 0
\end{array}\right]\otimes\left[\begin{array}{cc}
\gamma_{B} & 0\\
0 & 1
\end{array}\right],
\end{equation}
\begin{equation}
K_{4}=\left[\begin{array}{cc}
0 & 0\\
\omega_{A} & 0
\end{array}\right]\otimes\left[\begin{array}{cc}
0 & 0\\
\omega_{B} & 0
\end{array}\right],\label{K4}
\end{equation}
where $\gamma_{A}(t)$ and $\gamma_{B}(t)$ are two functions
describing the decay process (details can be found in Refs. \cite{Yu-Eberly04,YuPRB,Liu}). 
Given $\gamma_{A}(t)$  and $\gamma_{B}(t)$, $\omega_{A}(t)$ and $\omega_{B}(t)$ can be
determined as $\omega_{A}(t)=\sqrt{1-\gamma_{A}^{2}}$, $\omega_{B}(t)=\sqrt{1-\gamma_{B}^{2}}$.
Note that our scheme is applicable to arbitrary quantum channels. For example, in the finite temperature case \cite{Wang}, there exists two invertible Kraus operators, thus allowing our scheme to be pertinent. 
In the scheme implemented by Kim's group ($T=0$) \cite{Kim}, the successful probability of restoring the initial state $|\psi(0)\rangle=\alpha|00\rangle+\beta|11\rangle$ is
\begin{equation}
P_{WM}=\gamma_{A}^{2}\gamma_{B}^{2}\bar{p}_{1}\bar{p}_{2}[1+|\beta|^{2}(\bar{p}_{2}\omega_{B}^{2}+\bar{p}_{1}\omega_{A}^{2}+\bar{p}_{1}\bar{p}_{2}\omega_{A}^{2}\omega_{B}^{2})].\label{PWM}
\end{equation}
where $\bar{p}_1$ and $\bar{p}_2$ represent the measurement strengths of the weak measurement before sending the state into the dissipative channel.

On the other hand, in our EA-WMR scheme,  given the four Kraus
operators in Eq.~(\ref{K1}-\ref{K4}), it is clear that only $K_{1}$
is invertible, while other Kraus operators are singular. The smallest 
eigenvalue of $K_{1}K_{1}^{\dagger}$ is  $\gamma_{A}^{2}\gamma_{B}^{2}$,
so that
\begin{equation}
N_{1}=\gamma_{A}\gamma_{B},\; N_{i}=0\;(i=2,3,4).
\end{equation}
The final probability is
\begin{equation}
P_{EW}=\gamma_{A}^{2}\gamma_{B}^{2}.
\end{equation}
If we define a ratio as $\frac{P_{WM}}{P_{EW}}$, it is straightforward
to obtain 
\begin{equation}
\frac{P_{WM}}{P_{EW}}=\bar{p}_{1}\bar{p}_{2}[1+|\beta|^{2}(\bar{p}_{2}\omega_{B}^{2}+\bar{p}_{1}\omega_{A}^{2}+\bar{p}_{1}\bar{p}_{2}\omega_{A}^{2}\omega_{B}^{2})].
\end{equation}
For the short-time limit, and assuming the symmetric case, $\omega_{A}^{2}=\omega_{B}^{2}\approx0$,
we always have
\begin{equation}
\frac{P_{WM}}{P_{EW}}\ll1,
\end{equation}
since $\bar{p}_{1}$ and $\bar{p}_{2}$ must be chosen close to zero
to ensure high fidelity \cite{Kim}.

\begin{figure}
\noindent \begin{centering}
\includegraphics[width=1.0\columnwidth]{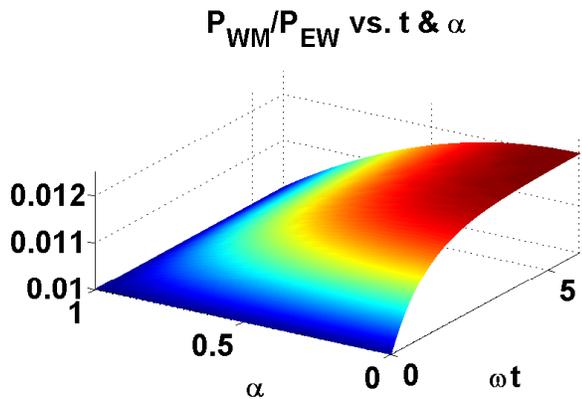}
\par\end{centering}

\caption{(Color online) The ratio $\frac{P_{WM}}{P_{EW}}$ at different times  $t$ with different
initial states ($1 \geq \alpha \geq 0$) . The initial states are given by
 $|\psi(0)\rangle=\alpha|00\rangle+\beta|11\rangle$. It is 
clear that this ratio is always smaller than $1$, which
means that the environment-assisted error correction scheme always give a higher successful probability.}

\label{ratio}
\end{figure}

For the long-time limit, $\omega_{A}^{2}=\omega_{B}^{2}\approx1$,  we
also have
\begin{equation}
\frac{P_{WM}}{P_{EW}}\ll1,
\end{equation}
since every term in the ratio is at least a second order function
of $\bar{p}_{1}$ and $\bar{p}_{2}$. If $\bar{p}_{1}$ and $\bar{p}_{2}$
are small, the ratio must be greatly smaller than $1$. The
detailed discussion is given in our numerical simulation. The numerical result
in Fig.~\ref{ratio} confirms the theoretical analyzation above. In
Fig.~\ref{ratio}, the ratio $\frac{P_{WM}}{P_{EW}}$ is always less
than one for an arbitrary initial state at any point in time.

\noindent 

Finally, it is worth noting that the EA-WMR always give unit fidelity. The fidelity of WMR scheme is 
highly dependent on the choice
of measurement strength $p_{1}$ and $p_{2}$. In general, the fidelity 
can be close to $1$, but it can never reach $1$. However, we should mention that the WMR has used less information
(resource) than the scheme proposed here.  Clearly, the enhanced successful probability and fidelity of the EA-WMR is due to the 
additional information obtained  from the environment measurement.

\section{\label{sec:V} Optimization of measurement basis}

As it is shown in Eq.~(\ref{PEW}) and Eq.~(\ref{eqN}), the successful
probability of the newly proposed EA-WMR scheme depends on the choice
of the Kraus decomposition. In the example shown in the last section,
only one Kraus operator is non-singular (invertible), which means
only one of the trajectories can be restored to the initial state,
and the other outcomes from the measurement of the environment must
be discarded. Therefore, it is of interest to consider how to choose
the Kraus decomposition to maximize the the successful probability.
In general this optimization issue can be very complex.  However, in what follows, we will try to
use a simple model to exemplify how to approach this interesting problem.

\begin{figure}
\noindent \begin{centering}
\includegraphics[width=1\columnwidth]{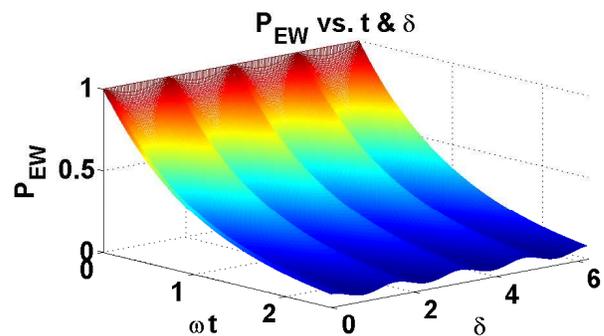}
\par\end{centering}

\caption{(Color online) Successful probability for all the possible Kraus decompositions.
The parameter $\delta$ is in the unit of radians.}
\label{optimization}
\end{figure}

To begin with, we consider a simple model consisting of one qubit in the dissipative
channel. There is a well-known Kraus decomposition for this model as \cite{QCQI,YuPRB,Liu}
\begin{equation}
\rho(t)=K_{1}\rho(0)K_{1}^{\dagger}+K_{2}\rho(0)K_{2}^{\dagger},
\end{equation}
where 
\begin{equation}
K_{1}=\left[\begin{array}{cc}
\gamma(t) & 0\\
0 & 1
\end{array}\right],\; K_{2}=\left[\begin{array}{cc}
0 & 0\\
\omega(t) & 0
\end{array}\right].\label{Kraus1qu}
\end{equation}
But the Kraus decomposition is not unique. In fact, any linear combination
of $K_{1}$and $K_{2}$ in the following form will also be a valid
set of Kraus operators:
\begin{equation}
L_{n}={\textstyle \sum_{m}}V_{nm}K_{m}.\label{eq:2-1}
\end{equation}
where $V_{nm}$ is a unitary matrix.  In general, an arbitrary two by
two unitary matrix can be written as
\begin{eqnarray}
V & = & \left[\begin{array}{cc}
e^{i(\alpha-\beta-\gamma)}\cos\delta & -e^{i(\alpha-\beta+\gamma)}\sin\delta\\
e^{i(\alpha+\beta-\gamma)}\sin\delta & e^{i(\alpha+\beta+\gamma)}\cos\delta
\end{array}\right]\nonumber \\
 & = & e^{i\alpha}\left[\begin{array}{cc}
e^{-i\beta} & 0\\
0 & e^{i\beta}
\end{array}\right]\left[\begin{array}{cc}
\cos\delta & -\sin\delta\\
\sin\delta & \cos\delta
\end{array}\right]\left[\begin{array}{cc}
e^{-i\gamma} & 0\\
0 & e^{i\gamma}
\end{array}\right].
\end{eqnarray}
However, we could prove that the final successful probability only
depends on the parameter $\delta$ of the transformation matrix $V$.
According to Eq.~(\ref{PEW}), since the probability only depends
on the eigenvalues of $L_{n}L_{n}^{\dagger}$, for example,
\begin{eqnarray}
L_{1}L_{1}^{\dagger} & = & |V_{11}|^{2}K_{1}K_{1}^{\dagger}+|V_{12}|^{2}K_{2}K_{2}^{\dagger}\nonumber \\
 &  & +V_{11}V_{12}^{*}K_{1}K_{2}^{\dagger}+V_{12}V_{11}^{*}K_{2}K_{1}^{\dagger}.
\end{eqnarray}
Solving the Secular equation $|L_{1}L_{1}^{\dagger}-\lambda I|=0$,
we will find the phase factors $\alpha$, $\beta$, $\gamma$ are
all canceled, where only the parameter $\delta$ survives. Therefore, we
only need to consider the single-parameter transformation matrix
\begin{equation}
V=\left[\begin{array}{cc}
\cos\delta & -\sin\delta\\
\sin\delta & \cos\delta
\end{array}\right].
\end{equation}
We have performed extensive numerical summations to test unitary transformations. The result is plotted in Fig.~\ref{optimization}. From the figure, we find the highest successful probability appears at
$\delta=n\pi/2$, where $n$ is an integer. Interestingly, it turns out that the choice (\ref{Kraus1qu}) actually gives the highest successful probability. Notably, the
measurement basis for the original Kraus decomposition (\ref{Kraus1qu}) is also a natural one. 
In this decomposition, we only need to perform the parity measurement \cite{parity1,parity2} to distinguish
two possible trajectories $K_{1}\rho(0)K_{1}^{\dagger}$ and $K_{2}\rho(0)K_{2}^{\dagger}$. 

\section{\label{sec:VI}Conclusion}

In this paper, we have proposed a new scheme to recover quantum states decohered by a noisy environment. 
The new scheme has generalized the EAEC scheme to the cases involving non-RU decompositions and 
weak measurement operations.  We show though a dissipative model how to use a weak measurement to restore 
the initial quantum state even when the Kraus operators are not of an RU type. It should be
noted that the new scheme we designed is a probabilistic process, rather than a deterministic one. 
In addition, we illustrate the new scheme in a particular example and compare its successful probability against that of the WMR scheme. We have shown that, with the aid of environment measurement, 
the successful probability can be enhanced. Finally, we also discuss the possibility
of improving the successful probability by choosing different Kraus
decompositions. Our future research will be focused on the application of our new schemes to multiple qubit systems.

\begin{acknowledgments}
We thank J. H. Eberly and W. T. Strunz for their useful discussions. We acknowledge
grant support from DOD/AF/AFOSR No. FA9550-12-1-0001. 
\end{acknowledgments}



\begin{thebibliography}{10}
\bibitem{Normalization}In the case that $K_n$ is singular, the smallest eigenvalue of $K_nK_n^\dagger$ is zero. Coincidently, the probability for restoration when in that case is just zero. So, the equation cover all the possible cases including the singular case.
\end{thebibliography}
\end{document}